\documentclass[useAMS,usenatbib]{mn2e} 
\usepackage{amsmath}
\usepackage{graphicx} 
\usepackage{aas_macros}
\usepackage[usenames]{color} 

\title[Oscillations in stellar superflares]
{Oscillations in stellar superflares}
\author[L.A. Balona, A.-M. Broomhall, A. Kosovichev, V. M. Nakariakov,
C.E. Pugh, T. Van Doorsselaere]
{L.A. Balona$^1$, A.-M. Broomhall$^{2,3}$, A. Kosovichev$^4$, V. M.
Nakariakov$^{2,5}$, \newauthor{C.E. Pugh$^2$, T. Van Doorsselaere$^6$}\\
$^1$South African Astronomical Observatory, P.O. Box 9, Observatory 7935,
South Africa\\
$^2$Centre for Fusion, Space and Astrophysics, Department of Physics,
University of Warwick, CV4 7AL, UK\\
$^3$Institute of Advanced Studies, University of Warwick, CV4 7HS,  UK\\
$^4$New Jersey Institute of Technology, Newark, NJ 07103, USA\\
$^5$Central Astronomical Observatory at Pulkovo of the Russian Academy
of Sciences, St Petersburg 196140, Russia\\
$^6$Centre for mathematical Plasma Astrophysics, Department of Mathematics, 
KU Leuven,\\
Celestijnenlaan 200B bus 2400, B-3001 Leuven, Belgium
}

\date{}

\def\LaTeX{L\kern-.36em\raise.3ex\hbox{a}\kern-.15em 
 T\kern-.1667em\lower.7ex\hbox{E}\kern-.125emX} 
 
\begin{document}

\maketitle 
 
\begin{abstract}
Two different mechanisms may act to induce quasi-periodic pulsations (QPP) 
in whole-disk observations of stellar flares.  One mechanism may be
magneto-hydromagnetic (MHD) forces and other processes acting on flare
loops as seen in the Sun.  The other mechanism may be forced local acoustic 
oscillations due to the high-energy particle impulse generated by the
flare (known as `sunquakes' in the Sun).  We analyze short-cadence 
{\it Kepler} data of 257 flares in 75 stars to search for QPP in the 
flare decay branch or post-flare oscillations which may be attributed to 
either of these two mechanisms.  About 18 percent of stellar flares show a
distinct bump in the flare decay branch of unknown origin.  The bump
does not seem to be a highly-damped global oscillation because the periods
of the bumps derived from wavelet analysis do not correlate with any stellar 
parameter.  We detected damped oscillations covering several cycles (QPP), 
in seven flares on five stars.  The periods of these oscillations also do 
not correlate with any stellar parameter, suggesting that these may be a 
due to flare loop oscillations.  We searched for forced global oscillations 
which might result after a strong flare.  To this end, we investigated the 
behaviour of the amplitudes of solar-like oscillations in eight stars before 
and after a flare.  However, no clear amplitude change could be detected.  
We also analyzed the amplitudes of the self-excited pulsations in two 
$\delta$~Scuti stars and one $\gamma$~Doradus star before and after a flare.  
Again, no clear amplitude changes were found.  Our conclusions are that a 
new process needs to be found to explain the high incidence of bumps in 
stellar flare light curves, that flare loop oscillations may have been 
detected in a few stars and that no conclusive evidence exists as yet 
for flare induced global acoustic oscillations (starquakes).
\end{abstract} 
 

\section{Introduction}

Photometric observations from the {\em Kepler} spacecraft have revealed
flares in a considerable number of stars.  These occur not only in cool M
dwarfs, but in K, G, F and A stars as well.  The numbers of B stars observed
by {\it Kepler} are too few to allow detection of flares.  Because even the
weakest visible stellar flares are thousands of times more energetic than a large
solar flare, these are often called ``superflares''.  \citet{Walkowicz2011} 
were the first to identify 373 flare stars out of $\approx 23000$ cool dwarfs 
in the  {\it Kepler} field.  Subsequently, \citet{Maehara2012} discovered 148 
solar-type stars with superflares.  More recently, \citet{Shibayama2013} 
found superflares on 279 G dwarfs.  \citet{Balona2012c} and 
\citet{Balona2013c} discovered several A stars with superflares which
cannot be attributed to flares in a cool companion. It appears that A stars 
may have spots and flares in spite of the lack of significant convection.  
In fact, the incidence of flares on A stars is not much lower than in G
and F stars \citep{Balona2015a}.  In these hot stars the magnetic field may
be a result of the Tayler instability in a differentially-rotating star
\citep{Spruit2002, Mullan2005}.

Whereas solar flares emit predominantly in several discrete chromospheric
and UV lines of highly-ionized elements, the optical flux in stellar 
flares appears to be distributed much like the continuum of an A or B star 
\citep{Kowalski2013}.  White-light emission in solar flares is difficult to
detect, though it is possible that it might contribute a large fraction of
the total flare radiation \citep{Kretzschmar2011}.  Even the strongest
solar flares are barely detected in space observations of total solar
irradiance and the Sun would not be detected as a flare star by {\it Kepler}.
It is thought that solar flares arise from the energy released by magnetic 
reconnection.  Although superflares are typically $10^6$ times more energetic 
than large solar flares, it is possible that the magnetic reconnection model 
may still apply \citep{Shibata2013}.

Nearly all {\em Kepler} photometry has been obtained with 30-min
long-cadence (LC) exposures, which means that only flares which have a long 
duration can be detected.  A few thousand stars were observed with 
1-min short-cadence (SC) exposures, though only for a relatively short time.  
Whereas LC data are nearly continuous over the four-year time span of {\it Kepler} 
observations, the SC data cover typically only a few months for any star.  The 
advantage of SC data is that it allows flares of short duration to be detected.  
\citet{Balona2015a} discovered 3140 flares in 209 stars of all spectral types 
observed in SC mode.  In this paper the flare in a particular star is
identified by its sequence number.  For example 002300039-028 is the 28-th
flare in KIC\,2300039 listed in the catalogue of \citet{Balona2015a}. 

We know that quasi-periodic pulsations (QPP) of coronal loops  occur in some 
solar flares \citep{Nakariakov2009, Nakariakov2010}.  The mechanism
giving rise to QPP is not fully understood at present.  One possibility
pointed out by \citet{McLaughlin2012} is that of oscillatory reconnection.
The behaviour of oscillatory reconnection is similar to a damped harmonic 
oscillator and may play a role in generating QPP. 

Analysis of QPP can provide information on the flare coronal environment and 
magnetic field strength by the use of seismology \citep{Kupriyanova2013}.  
However, solar flare QPP are only observed in H$\alpha$, extreme UV lines and 
in radio and X-ray and gamma-ray observations.  There are no recorded 
observations of QPP in white-light solar flares.  The reported optical 
observations of QPP in stellar flares \citep{Rodono1974, Zhilyaev2000, 
Mathioudakis2006, Contadakis2010, Qian2012, Anfinogentov2013}, which have 
periods ranging from a few seconds to tens of minutes, are whole-disk 
essentially white-light observations.  They cannot therefore be directly 
compared with solar flare QPP.  It is also far from certain if white-light 
stellar flares are generated in the same way as solar flares.

Solar and stellar QPP have previously been used to provide estimates for
several flare parameters.  An example of QPP with large amplitude and duration 
in a solar flare observed in X-ray and microwave radio bursts in described
by \citet{Kane1983}.  More recently, \citet{VanDoorsselaere2011}
used X-ray observations of two oscillation modes in a single solar flare to
estimate the  plasma-beta and the density contrast of the flaring loop.
The wave mode number was also estimated from the observed periods.
\citet{Anfinogentov2013} analyzed the oscillatory signal in the decay phase
of the U-band light curve of a flare in the dM4.5e star YZ~CMi.
The observational signature is typical of the longitudinal oscillations
observed in solar flares at extreme ultraviolet and radio wavelengths and
is associated with standing slow magnetoacoustic waves.  They therefore
suggest that the QPP in this stellar superflare may be of a similar nature 
to solar QPP.  A well pronounced QPP has been reported during an very
energetic flare on the RS~CVn binary II\,Peg \citep{Mathioudakis2003}.  The 
QPP has a long period leading to the peak. 

The dynamic impact in the photosphere caused by a solar flare is called a
``sunquake''.  The resulting helioseismic waves are observed as expanding 
circular ripples on the solar surface, which can be detected in Dopplergrams
and as a characteristic ridge in time-distance diagrams \citep{Kosovichev1998, 
Kosovichev2006}, or by calculating the integrated acoustic emission 
\citep{Donea1999, Donea2005}.   These flare-excited oscillations 
are mostly local seismic waves.  While the theory predicts that global 
acoustic waves should also be excited, their amplitudes are thought to be 
significantly lower than the amplitudes of stochastically excited oscillations
\citep{Kosovichev2009}.  Most of the emitted acoustic energy of sunquakes is
above the acoustic cut-off frequency of the Sun.

\begin{figure*}
\centering
\includegraphics{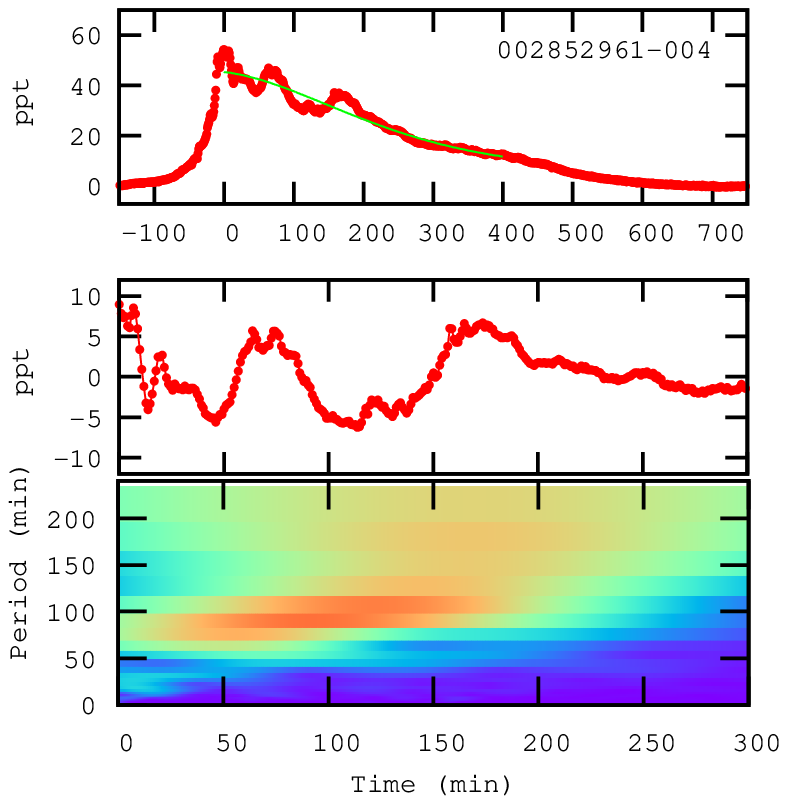}
\includegraphics{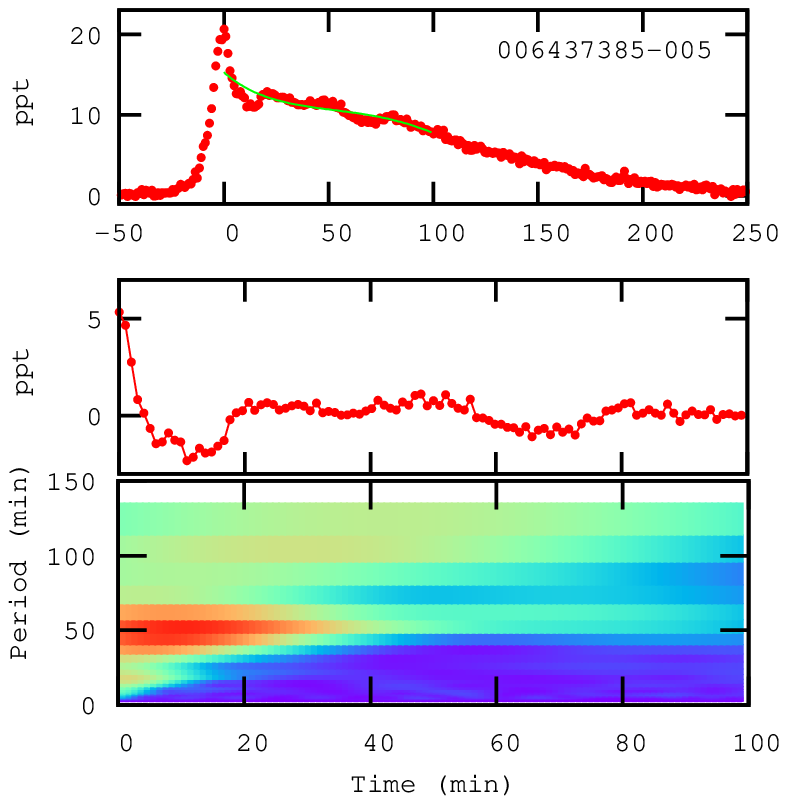}
\includegraphics{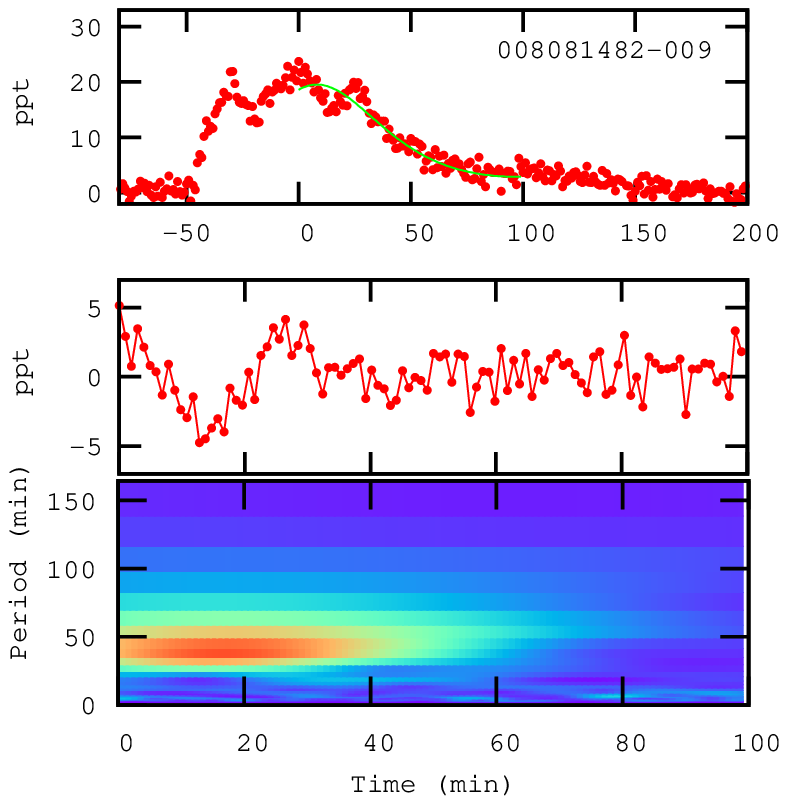}
\includegraphics{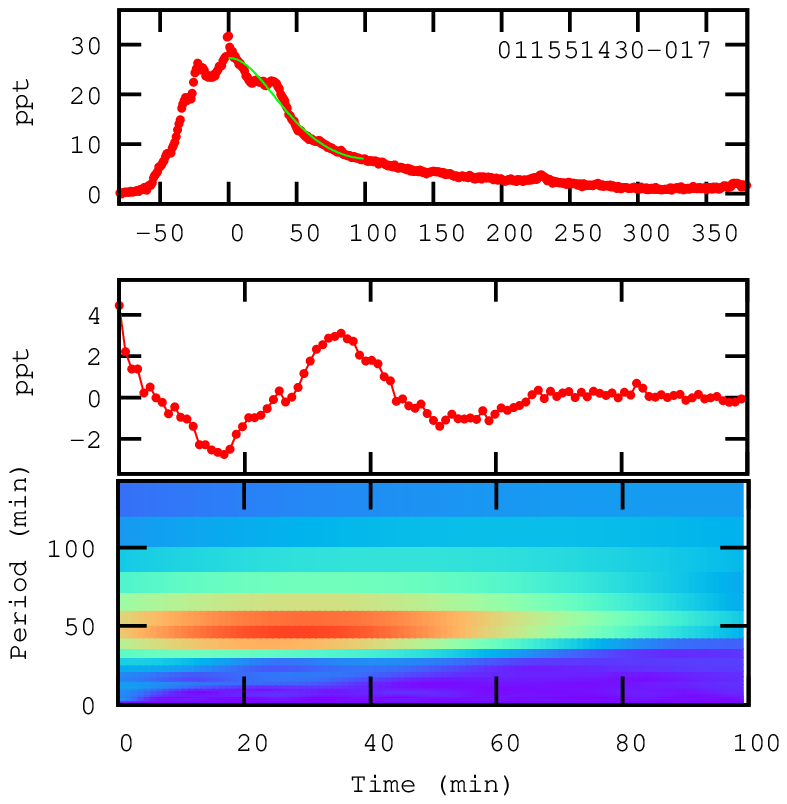}
\caption{Examples of four flares showing bumps on the decay branch.  The top
panel shows the flare light curve and fitted polynomial.  The middle panel
shows the decay branch with polynomial removed, with time measured from the
time of maximum flare intensity.  The bottom panel shows the wavelet 
spectrum as a function of time relative to the time of maximum flare
intensity.  The intensity is measured in parts per thousand (ppt).  The KIC
number and the flare identifier from the catalogue of \citet{Balona2015a} is
shown.}
\label{long}
\end{figure*}

\citet{Karoff2008a} concluded that global high-frequency solar acoustic 
waves have larger amplitudes after some solar flares, a finding confirmed 
by \citet{Kumar2010}.  However, \citet{Richardson2012} found that a decrease 
in acoustic power after a solar flare is just as likely as an increase.  This 
is perhaps not surprising because the effect of the impulse on a global seismic 
mode depends on the location and time of the impulse.  The effect of the impulse
is therefore expected to increase the amplitude of some modes and decrease
the amplitude in other modes.  The result should be a higher variation in
amplitude distribution after a flare.

In the case of the vastly more powerful stellar superflares, the impact on the 
star may be much greater and it is conceivable that these forced oscillations 
may be observed.  This idea has been discussed by \citet{Karoff2014} and
\citet{Kosovichev2014} in {\it Kepler} photometry of solar-like stars.
Karoff found no significant enhancement of the energy in the post-flare 
acoustic spectra relative to the pre-flare energy.  However, a larger 
variability between the energy in the high-frequency part of the post- and 
pre-flare acoustic spectra was found compared to spectra taken at random 
times.  This may be a result of the increased dispersion in acoustic energy 
discussed above.

In spite of the very large differences in energy and optical emission 
between solar and stellar flares, the possibility exists that the physical 
mechanisms for QPP could be very similar. In this paper we investigate
QPP in {\it Kepler} superflares because this could provide additional clues 
to the nature of stellar flares.  We also investigate the possible occurrence 
of starquakes as a result of a superflare and search for excitation of
global acoustic oscillation modes in the periodogram.  These investigations 
are only possible due to the superb precision of the  {\it Kepler} data, a 
unique resource which is unlikely to be equaled for many years to come.

\section{Data and analysis technique}

Our analysis is based on the 3140 flares in 290 stars observed in SC mode as
described in \citet{Balona2015a}.  Out of these 3140 flares, we selected
257 flares in 75 stars with sufficiently high signal-to-noise (S/N) to be 
suitable for detecting possible flare oscillations.

\setlength{\tabcolsep}{4pt}

\begin{table*}  
\begin{center}  
\caption{Physical parameters for stars in which one or more bumps are
visible in the flare light curve.  The second column, $N_f$, is an identifier
for the flare, which is the $N_f$-th flare for the particular star in the 
catalogue of \citet{Balona2015a}. The {\em Kepler} magnitude, $Kp$,  effective 
temperature, $T_{\rm eff}$ (K) and surface gravity, $\log g$, is taken from the
{\em Kepler Input Catalogue} (KIC).  The luminosity, $L/L_\odot$ is derived 
from the effective temperature and radius listed in the KIC.  The average 
number of flares per day, $N/d$, and the total number of observed flares, 
$N$, are shown.  The rotation period, $P_{\rm rot}$ (d), are from 
\citet{Balona2015a}. $P_{\rm max}$ is the expected period of solar-like 
oscillations (min).  The last column is the derived period from wavelet 
analysis.}
\label{starinfo}
\vspace{2mm}  
\begin{tabular}{rrrrrrrrrrl}
\hline  
\multicolumn{1}{c}{KIC}             &
\multicolumn{1}{c}{$N_f$}             &
\multicolumn{1}{c}{Kp}              &
\multicolumn{1}{c}{$T_{\rm eff}$}   &  
\multicolumn{1}{c}{$\log g$}        &  
\multicolumn{1}{c}{$\log L/L_\odot$}&  
\multicolumn{1}{c}{$N/d$}        &  
\multicolumn{1}{c}{$N$}           &  
\multicolumn{1}{c}{$P_{\rm rot}$}   &
\multicolumn{1}{c}{$P_{\rm max}$}   &
\multicolumn{1}{l}{$P_{\rm bump}$}  \\
\multicolumn{1}{c}{}                &
\multicolumn{1}{c}{}                &
\multicolumn{1}{c}{(K)}               &  
\multicolumn{1}{c}{(dex)}             &  
\multicolumn{1}{c}{}                &  
\multicolumn{1}{c}{}                &  
\multicolumn{1}{c}{}           &  
\multicolumn{1}{c}{(d)}   &
\multicolumn{1}{l}{(min)}  &
\multicolumn{1}{l}{(min)}  \\
\hline
\\
   2300039 & 28 & 15.42 & 3644 &   4.294 &  -0.9049 &   0.710 &   65 &   1.707 &    5.9 &  4.8 \\
   2852961 &  1 & 10.14 & 4722 &   2.877 &   1.3730 &   0.201 &    5 &   8.8:  &  175.9 & 80.6 \\
           &  3 &       &      &         &          &         &      &         &        & 68.0 \\
           &  4 &       &      &         &          &         &      &         &        & 97.0 \\
   4671547 & 59 & 11.29 & 4059 &   4.653 &  -1.1032 &   2.453 &   64 &   8.138 &    2.7 & 12.5 \\
           & 62 &       &      &         &          &         &      &         &        &  7.0 \\
   5475645 &  3 & 11.20 & 5336 &   4.654 &  -0.3709 &   0.059 &    6 &   7.452 &    3.1 &  9.2 \\
   5733906 &  2 & 11.83 & 5241 &   3.688 &   0.6933 &   0.238 &    7 &   0.719 &   28.6 &  5.5 \\
   5952403 &  2 &  6.96 & 5037 &   3.001 &   1.5178 &   0.012 &    2 &  45.28: &  136.5 &110.2 \\
   6205460 &  4 & 12.74 & 5242 &   3.677 &   0.7067 &   0.111 &   16 &   3.717 &   29.4 &  8.7 \\
           &  5 &       &      &         &          &         &      &         &        & 12.7 \\
           & 12 &       &      &         &          &         &      &         &        & 57.5 \\
   6437385 &  5 & 11.53 & 5401 &   3.713 &   0.7194 &   0.205 &   18 &  13.672 &   27.4 & 55.0 \\
           &  7 &       &      &         &          &         &      &         &        & 86.0 \\
   6548447 &  2 & 12.88 & 5031 &   4.005 &   0.2505 &   0.093 &    8 &   9.409 &   13.5 & 60.0 \\
   7885570 & 23 & 11.67 & 5398 &   4.616 &  -0.3033 &   0.167 &   40 &   1.730 &    3.4 & 38.0 \\
           & 27 &       &      &         &          &         &      &         &        & 40.0 \\
   7940533 &  4 & 12.86 & 5326 &   4.555 &  -0.2665 &   0.179 &   21 &   3.826 &    3.9 & 42.8 \\
           &  5 &       &      &         &          &         &      &         &        & 38.0 \\
   8081482 &  6 & 14.56 & 5522 &   4.333 &   0.0524 &   0.109 &   28 &   2.819 &    6.7 & 66.0 \\
           &  9 &       &      &         &          &         &      &         &        & 40.9 \\
   8226464 &  4 & 11.46 & 5754 &   4.053 &   0.4475 &   0.526 &   15 &   3.101 &   12.9 & 53.5 \\
           &  8 &       &      &         &          &         &      &         &        & 35.0 \\
   8608490 & 10 & 14.77 & 4897 &   3.955 &   0.2591 &   0.032 &   10 &   1.083 &   15.0 & 76.2 \\
   9349698 & 32 & 12.91 & 4911 &   4.537 &  -0.4318 &   1.861 &  136 &   1.359 &    3.9 & 51.6 \\
   9576197 &  7 & 14.64 & 5082 &   4.551 &  -0.3651 &   0.164 &   12 &   9.096 &    3.9 & 68.3 \\
   9641031 & 10 &  9.17 & 5867 &   4.295 &   0.2134 &   0.145 &   84 &   2.156 &    7.5 & 11.9 \\
           & 35 &       &      &         &          &         &      &         &        & 21.8 \\
   9655129 &  6 & 13.80 & 5140 &   4.431 &  -0.2067 &   0.113 &   20 &   2.750 &    5.1 & 40.9 \\
           &  7 &       &      &         &          &         &      &         &        & 19.8 \\
   9833666 &  7 &  9.68 & 5411 &   3.735 &   0.6977 &   0.164 &    9 &  10.341 &   26.1 & 42.6 \\
           &  9 &       &      &         &          &         &      &         &        & 40.0 \\
  10063343 & 30 & 13.16 & 3976 &   4.433 &  -0.8685 &   1.536 &   46 &   0.333 &    4.5 &  3.1 \\
  10976930 &  1 & 11.28 & 5934 &   3.644 &   0.9735 &   0.069 &    2 &   2.054 &   33.7 & 29.3 \\
  11445774 &  2 & 11.91 & 6108 &   4.328 &   0.2577 &   0.026 &    6 &   1.744 &    7.1 & 35.0 \\
  11551430 &  7 & 10.69 & 5335 &   3.729 &   0.6779 &   0.791 &  185 &   4.145 &   26.3 & 67.7 \\
           & 17 &       &      &         &          &         &      &         &        & 48.5 \\
           & 21 &       &      &         &          &         &      &         &        & 35.5 \\
           & 34 &       &      &         &          &         &      &         &        & 10.7 \\
           & 72 &       &      &         &          &         &      &         &        & 79.5 \\
           & 73 &       &      &         &          &         &      &         &        & 20.7 \\
           & 94 &       &      &         &          &         &      &         &        & 43.8 \\
  11610797 & 18 & 11.53 & 5865 &   4.464 &   0.0338 &   1.018 &   34 &   1.625 &    5.1 & 16.2 \\
  12156549 & 24 & 15.88 & 5541 &   4.378 &   0.0116 &   0.620 &  128 &   3.651 &    6.0 & 59.8 \\
\\
\hline
\end{tabular}
\end{center}
\end{table*}

\begin{figure}
\centering
\includegraphics{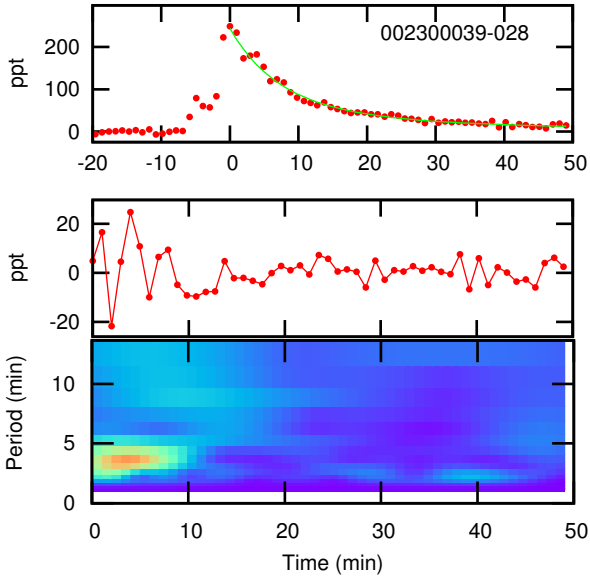}
\caption{The top panel shows the flare light curve and fitted spline curve.  
The second panel shows the decay branch with fitted curve removed.  The
third panel shows the wavelet spectrum.}
\label{fast1}
\end{figure}

\begin{figure}
\centering
\includegraphics{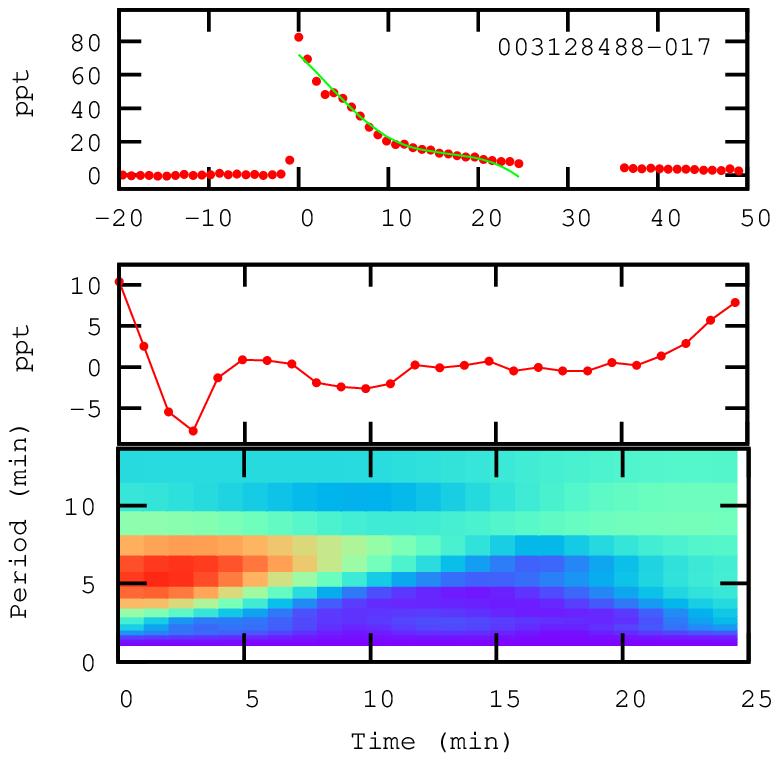}
\caption{The same as Fig.\,\ref{fast1}.}
\label{fast2}
\end{figure}

\begin{figure}
\centering
\includegraphics{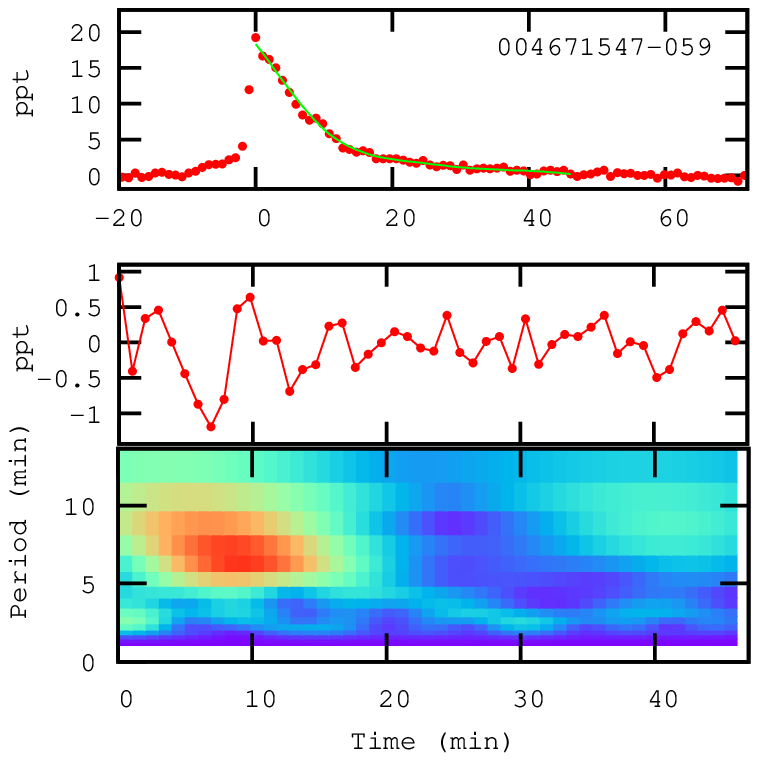}
\caption{The same as Fig.\,\ref{fast1}.}
\label{fast3}
\end{figure}

In order to enhance the visibility of possible oscillations it is necessary 
to remove the underlying flare decay.  We found that a polynomial of the form  
$$\log(y) = a_0 + a_1t + a_2t^2 + a_3t^3 \dots$$
generally provides a good fit to the flare decay intensity.  Here $t$ is the 
time measured from flare maximum and $y$ is the flare intensity.  In nearly 
every case, a cubic polynomial was used.  The polynomial was removed and the 
residuals plotted as a function  of time.  In about 18 percent of the {\it
Kepler} flares, one or more bumps are present in the decay branch.  A 
low-degree polynomial does not adequately remove these features.  In these 
cases, the large-scale structures can be removed by applying a suitable 
filter to the data, as described below.

Flare loop oscillations (QPP) are usually seen as damped sinusoidal oscillations
(there are however decay-less, multi-modal and wave-train regimes observed).
Wavelet analysis is an appropriate tool for detecting such a signal.  Wavelet 
analysis decomposes a time series into time/frequency space simultaneously.  
One gets information on both the amplitude of any periodic signals within 
the series, and how this amplitude varies with time.  This is accomplished by 
selecting an appropriate wavelet function that can be scaled and translated.   
The resulting wavelet transform is a representation of the signal at different 
time scales.  We use the Morlet wavelet (a sine wave multiplied by a Gaussian 
envelope) as the basis function in our analysis.  The results are shown in
a plot of instantaneous wave period and amplitude as a function of time.  This 
is conveniently accomplished by a greyscale representation of the amplitude
in a period--time diagram.

Forced global acoustic oscillations caused by a flare impulse are expected
to be seen several hours after the flare and with a decay time of perhaps a
few days.  For detecting such oscillations we use standard periodogram 
analysis applied to a data window with a length of a few days and starting a 
few hours after the flare.

\section{Flares with bumps in the light curve}  

Of 257 stellar flares selected because of their very high S/N, 47 flares
(i.e. 18\,percent) show obvious structures (i.e. bumps) in the flare decay 
light curve.  A further 14\,percent show a distinct change in the rate of
decay \citep{Balona2015a}.   Examples of flare light curves showing bumps are 
shown in Fig.\,\ref{long}.  One possible reason for such bumps is that two
flares occur by chance within the same small interval.  One can estimate the
chances of such an occurrence by measuring the mean flare rate during a
suitable interval.  The typical duration of a flare is about one hour.  The
star with the maximum flare rate produces, on average, one flare every
10\,hr or $x = 0.1$\,flare hr$^{-1}$.  The probability, $P$, that two flares 
will occur in the interval of one hour follows a Poisson distribution and is 
$P = \frac{x^2}{2!}e^{-x} = 0.005$.  This is an upper limit considering the 
fact that we have chosen the maximum flare rate.  In some stars (e.g.
KIC\,11551430) there are many flares with bumps.  The probability of such an
occurrence, assuming that what we are seeing is just a superposition of two
or more practically simultaneous flares, is negligible.  Hence there must be
a physical process involved in producing the bumps.  One such process could
be sympathetic flaring where one solar flare may trigger another flare
\citep{Moon2002}.

Both bumps and changes in rate of decay can be modeled as QPP with a rapid 
decay rate.  In Fig.\,\ref{long} the polynomial fit to the decaying branch, 
and the wavelet spectra are shown.  Wavelet analysis of 47 ``bump'' flares in 
30 stars indicates that significant power is present at a certain period.  If 
we are to make progress in understanding this relatively common phenomenon, 
the ``period'' derived in this may offer some clue as to the nature of the 
bumps.

If we assume that the bump is due to a highly-damped global acoustic mode, 
for example, one may expect the period to be correlated with some stellar 
parameter or combination of parameters.  Solar-like modes in stars are 
excited because their periods are similar to the typical turn-over period of 
a convective cell and below the acoustic cut-off frequency.  As a result, the 
frequency of maximum amplitude, $\nu_{\rm max}$ is related to the stellar 
parameters as follows \citep{Brown1991, Kjeldsen1995a}:
\begin{align*}
\nu_{\rm max} &\approx \nu_{{\rm max}\odot}\frac{M/M_\odot}{(R/R_\odot)^2
\sqrt{T_{\rm eff}/T_{{\rm eff}\odot}}},
\end{align*}
where  the solar value for the frequency of maximum amplitude is 
$\nu_{{\rm max}\odot} = 3120$\,$\mu$Hz, while  $M/M_\odot$,
$R/R_\odot$ and $T_{\rm eff}/T_{{\rm eff}\odot}$ is the stellar mass,
radius and effective temperature relative to the Sun.  

If, by analogy with sunquakes, the stellar flares excite mostly acoustic
modes with frequencies close to the acoustic cut-off frequency, one might
expect a correlation between the period and the acoustic cut-off frequency.
Since the acoustic cut-off frequency is proportional to $\nu_{\rm max}$,
one may expect a correlation between the bump period and $P_{\rm max} = 
1/\nu_{\rm max}$.  The value of $P_{\rm max}$, as calculated from $\log g$ 
and $T_{\rm eff}$, is shown in Table\,\ref{starinfo}.

\begin{figure}
\centering
\includegraphics{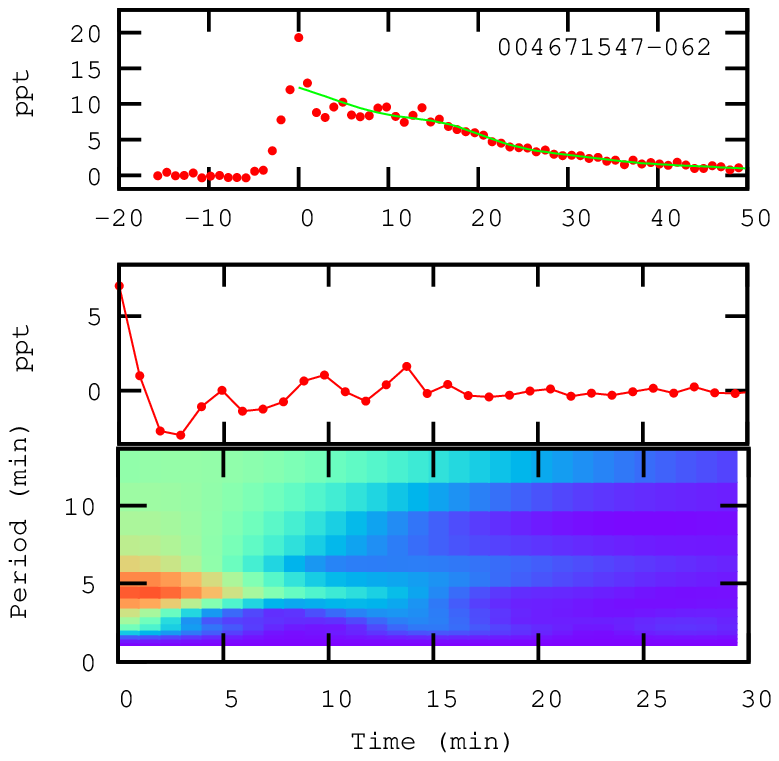}
\caption{The same as Fig.\,\ref{fast1}.}
\label{fast4}
\end{figure}

\begin{figure}
\centering
\includegraphics{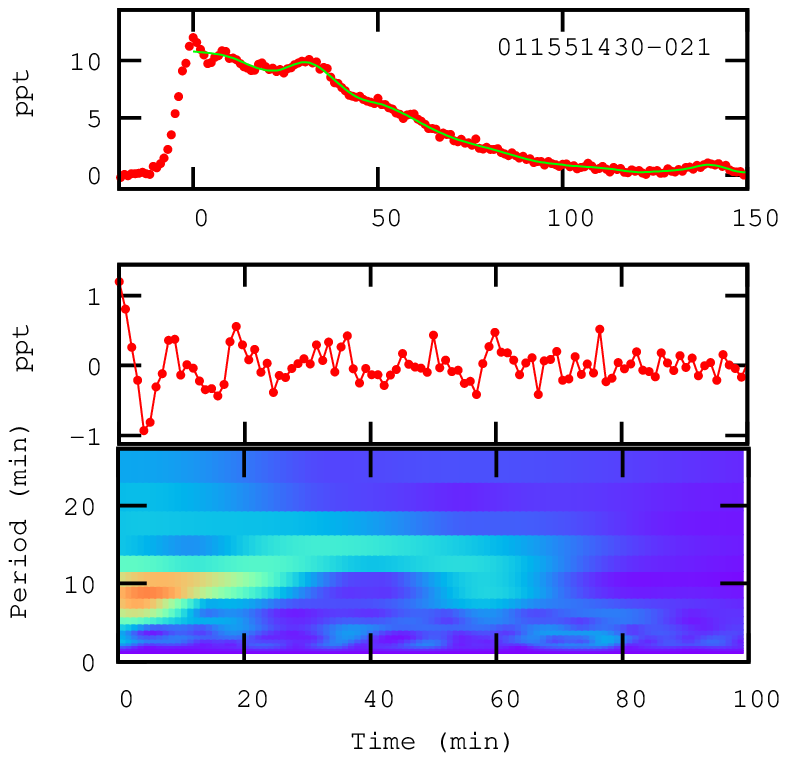}
\caption{The same as Fig.\,\ref{fast1}.}
\label{fast5}
\end{figure}

As shown in Table\,\ref{qppinfo}, the flare bump period ($P_{\rm bump}$) can vary 
widely even in the same star.  We can find no correlation between this period and 
$P_{\rm max}$.  In fact, no correlation between these timescales and any stellar 
parameter can be found.  It seems that the structures on the decaying branch of 
the flare, whatever their cause, cannot be attributed to highly-damped
impulsively excited global acoustic oscillations.  It is still possible that 
these may be highly-damped flare loop oscillations, but there is no evidence 
that can be used in support of this notion either.  However, the similarity of 
the observed decaying QPP in stellar superflares, and of the standing 
oscillations observed in hot coronal solar flare loops (e.g. 
\citealt{Wang2011, Kim2012}) indicates the possible similarity of the physical 
processes involved, despite differences in emission spectrum.

\section{Rapid oscillations}

A polynomial fit to the flare decay branch is no longer adequate to enhance the 
visibility of possible oscillations with periods of only a few minutes.  
We found a segmented spline fit to be adequate for this purpose.  In this
method, a number of evenly spaced points are selected in the decay branch of 
the flare light curve.  The mean intensity in the neighbourhood of each point 
is found.  A spline fit is calculated using the time and mean intensities at
these points and the fit removed from the data.  It is important to choose the 
interval between the segments carefully.  If the interval is too small, 
possible oscillations with periods longer than this interval will 
not be detected because less than one period will be sampled.  On the other 
hand, if the interval is too long, then the spline interpolation is no longer 
an adequate fit and the residuals may be contaminated by spurious long-period 
signals.  We found that a choice of 10--15\,min was appropriate for the 
interpolation sampling interval in most cases.  We have also tried
other methods for trend removal, such as temporal smoothing, but this did not
significantly influence the detected periods.

\begin{figure}
\centering
\includegraphics{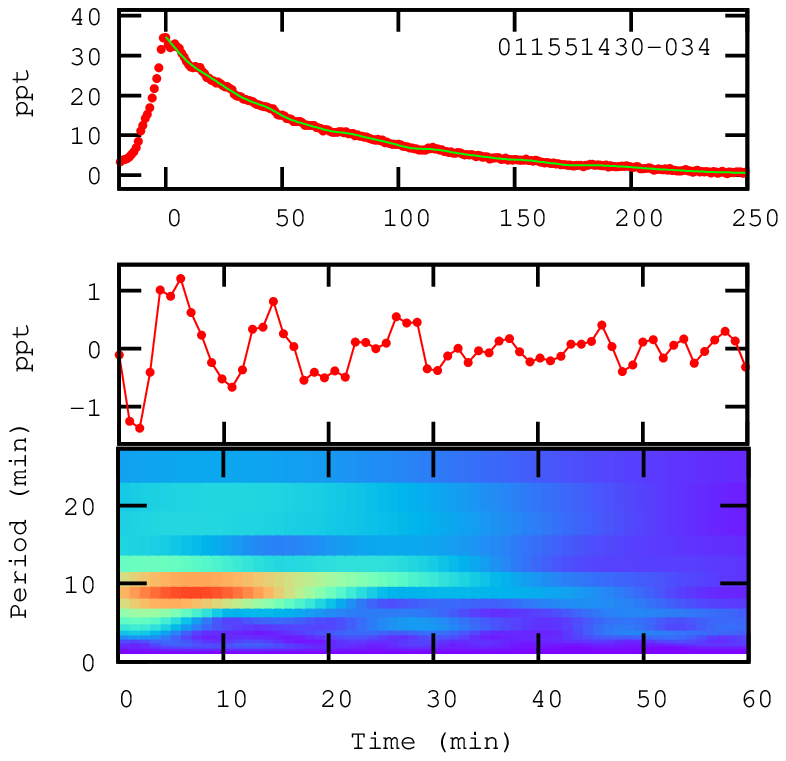}
\caption{The same as Fig.\,\ref{fast1}.}
\label{fast6}
\end{figure}

\begin{figure}
\centering
\includegraphics{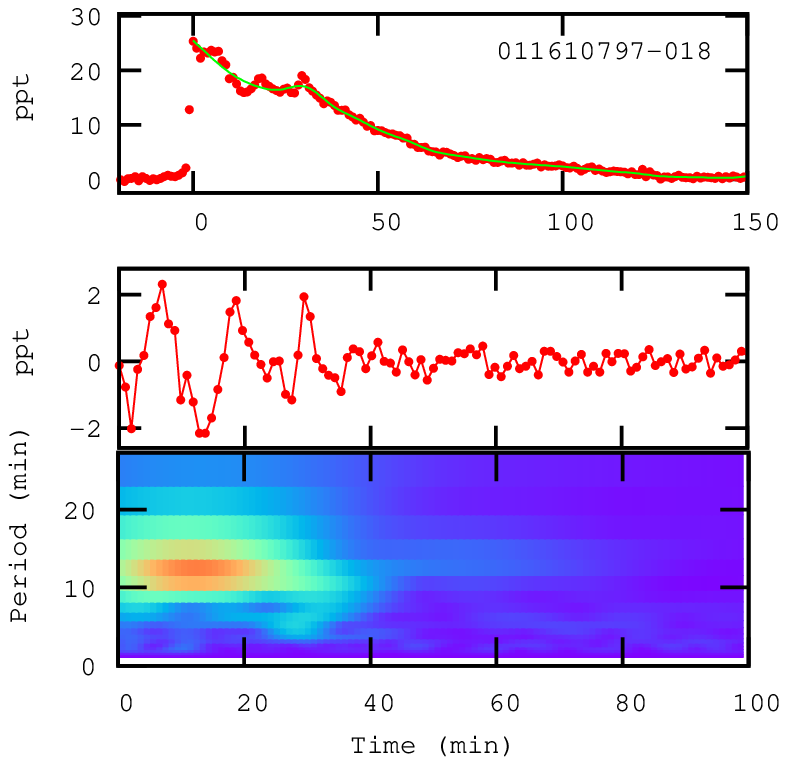}
\caption{The same as Fig.\,\ref{fast1}.}
\label{fast7}
\end{figure}

We carefully examined all 257 flares without finding any obvious QPP in the
majority of flares.  We did not find any cases where QPP begins before 
the time of flare maximum.  This may be due to the long sample time of
1\,min and the steep rising branch which which would make such a detection
difficult.  Evidence for damped oscillations after flare maximum is, however, 
present in the few flares shown in Figs.\,\ref{fast1}--\ref{fast7}.  The
stellar parameters for these stars are shown in Table\,\ref{qppinfo}.
Most of these flares also appear in Table\,\ref{starinfo}; other stars in 
this table do not show more than a single bump.  

\citet{Gruber2011} tested the apparent QPP of four bright solar flares 
observed in gamma rays using classical periodogram analysis, but found 
that these oscillations were not intrinsic to the flares.  Similarly, 
\citet{Vaughan2010} applied Bayesian statistics to apparent QPP in some 
Seyfert galaxies and also found that these were not significant.  The 
difficulty is that the underlying noise in the periodogram is not 
`white' (i.e., independent of frequency) but `red' (i.e. increases 
towards low frequencies).  This variation of noise with frequency
needs to be taken into account in any analysis of significance.  Because
white noise is usually assumed the significance of QPP has generally
been overestimated.  This is further illustrated by recent work by
\citet{Inglis2015} who investigated supposed QPP in a selection 
of solar flares from a variety of sources as well as QPP in some optical 
stellar flares.  They found that for all except one event tested, an 
explicit oscillation is not required in order to explain the observations.  
Instead, the flare signals are adequately described as a manifestation 
of a power law in the Fourier power spectrum, rather than a direct 
signature of oscillating components or structures.  

That does not mean that all QPP signals are spurious, but that great 
caution needs to be exercised in determining the significance of QPP in 
flares.  For this reason, we do not claim that the oscillating signals seen
in Figs.\,\ref{fast1}--\ref{fast7} are necessarily real.  We merely wish to
illustrate that QPP in the {\it Kepler} flares, if it exists, is not 
common.

\setlength{\tabcolsep}{4pt}

\begin{table}  
\begin{center}  
\caption{Physical parameters for stars in which a damped oscillation (QPP)
is visible in the flare light curve.   The second column, $N_f$, is an 
identifier for the flare. The {\em Kepler} magnitude, $Kp$, effective temperature, 
$T_{\rm eff}$ (K) and surface gravity, $\log g$, is taken from the 
{\em Kepler Input Catalogue} (KIC).    The average 
number of flares per day, $N/d$, and the total number of observed flares, 
$N$, are shown.  The variability type and the rotation period, 
$P_{\rm rot}$ (d), are from \citet{Balona2015a}.  The last column is 
the period (min) derived from wavelet analysis.}
\label{qppinfo}
\scriptsize
\vspace{2mm}  
\begin{tabular}{rrrrrrrrl}
\hline  
\multicolumn{1}{c}{KIC}             &
\multicolumn{1}{c}{$N_f$}             &
\multicolumn{1}{c}{Kp}              &
\multicolumn{1}{c}{$T_{\rm eff}$}   &  
\multicolumn{1}{c}{$\log g$}        &  
\multicolumn{1}{c}{$N/d$}        &  
\multicolumn{1}{c}{$N$}           &  
\multicolumn{1}{l}{$P_{\rm rot}$}   &
\multicolumn{1}{l}{$P_{\rm QPP}$}  \\
\multicolumn{1}{c}{}                &
\multicolumn{1}{c}{}                &
\multicolumn{1}{c}{(K)}               &  
\multicolumn{1}{c}{(dex)}             &  
\multicolumn{1}{c}{}                &  
\multicolumn{1}{c}{}                &  
\multicolumn{1}{c}{}           &  
\multicolumn{1}{c}{(d)}   &
\multicolumn{1}{c}{(min)}  \\
\hline
\\
   2300039 & 28 & 15.42 &  3644 &  4.294 & 0.710 &  65 &   1.707 &  4.8 \\
   3128488 & 17 & 11.66 &  4475 &  4.615 & 1.138 &  38 &   6.160 &  6.0 \\
   4671547 & 59 & 11.29 &  4059 &  4.653 & 2.453 &  64 &   8.138 &  6.0 \\
           & 62 &       &       &        &       &     &         &  5.0 \\
  11551430 & 21 & 10.69 &  5335 &  3.729 & 0.791 & 185 &   4.145 & 11.0 \\ 
           & 34 &       &       &        &       &     &         & 10.7 \\
  11610797 & 18 & 11.53 &  5865 &  4.464 & 1.018 &  34 &   1.625 & 14.0 \\
\\
\hline
\end{tabular}
\end{center}
\end{table}
\normalsize

While QPP is a useful diagnostic tool for flares on the Sun, it cannot be used 
for stellar flares without some underlying assumptions.  In the Sun
one can image the flare, so that the loop length is known.  Very often,
other parameters, such as the plasma temperature, can also be quite well
estimated.  For stellar flares, however, we only have the period and decay
time of the QPP, so the information that can be extracted is severely limited.
\citet{Zaitsev1982} and \citet{Roberts1984} showed that for a simple cylindrical magnetic flux tube,
several types of magneto-acoustic wave modes are possible:  the slow (acoustic)
mode, the fast kink and the fast sausage modes. These are all observed in
solar flux tubes and solar flare loops.  For a standing oscillation in 
a loop, the loop length $L$ is given by $L = jcP/2$, where $P$ is the 
oscillation period, $j$ the parallel mode number and $c$ the appropriate wave 
speed. The waves with the longest periods, which are those of interest to us 
due to the 1-min cadence of the {\it Kepler} data, are the slow modes.  For 
slow modes, $c$ is the tube speed, $c_t$ with
\begin{align*}
&\frac{1}{c_t^2} = \frac{1}{c_s^2} + \frac{1}{c_A^2},
\end{align*}
where $c_s$ is the sound speed and $c_A$ is the Alfv\'{e}n speed.  Since the
Alfv\'{e}n speed in the coronal loops is considerably larger than the sound 
speed, we assume $c \approx c_s$.  

The unknown values of $j$, $c$ and $L$ render the extraction of any meaningful 
physics impossible at this stage.  We also note that the above description 
applies to a low-density plasma environment as occurs in a typical solar
coronal flare loop.  Stellar flares emit in a continuum and the physical 
process is likely to be different from that of solar flares even if the energy
source, magnetic reconnection, is the same.  However, these parameters are contained in the 
characteristic parameters of the decay phase of the flare, the damping time of 
the oscillations, intensity of the flare, and other observables and, in 
principle, can be extracted from the data when a sufficiently detailed model 
becomes available.

\begin{figure}
\centering
\includegraphics{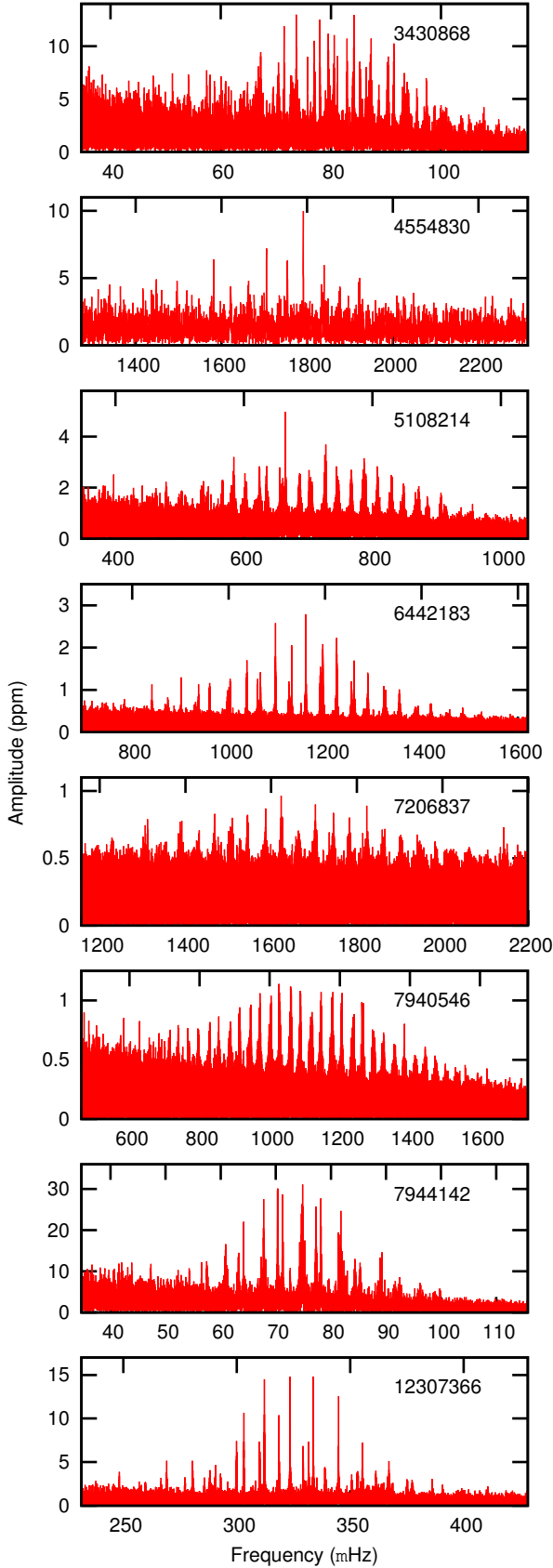}
\caption{Periodograms of flare stars with solar-like oscillations.}
\label{solfig}
\end{figure}

\section{Flares in stars with solar-like oscillations}

We know that some solar flares may excite local seismic disturbances
\citep{Kosovichev1998, Kosovichev2006}, with maximum power in waves of high
spherical harmonic degree, $l$.  These waves would not be visible in
whole-disk photometry because the net brightness change is canceled owing
to the small spatial wavelengths.  The power in waves with low $l$ which
might be visible is very low \citep{Kosovichev2005}.  The more powerful 
impulse provided by a stellar superflare may, however, lead to observable 
results (starquakes).  As described in the Introduction, the expected result 
is that the amplitudes of some global modes will grow while others will 
diminish after a flare.  It is therefore to be expected that the net 
result will be a larger variability of the energy in the global modes 
after the flare which can manifest itself as a larger variability in the
periodogram. 

The {\it Kepler} SC data is an ideal data set for studying possible
starquakes.  Among the 209 flare stars observed in SC mode, we can identify
eight stars (Table\,\ref{solar}) which clearly show the characteristic
Gaussian envelope of solar-like oscillations in the periodogram.  The
relevant portion of the periodograms are shown in Fig.\,\ref{solfig}.

\begin{table}  
\begin{center}  
\caption{Parameters for flare stars with solar-like oscillations.}
\label{solar}
\vspace{2mm}  
\begin{tabular}{rrrrrrrr}
\hline  
\multicolumn{1}{c}{KIC}             &
\multicolumn{1}{c}{Kp}              &
\multicolumn{1}{c}{$T_{\rm eff}$}   &  
\multicolumn{1}{c}{$\log g$}        &  
\multicolumn{1}{c}{$\log L/L_\odot$}&  
\multicolumn{1}{c}{$N/d$}           &  
\multicolumn{1}{c}{$N$}             &  
\multicolumn{1}{l}{$P_{\rm rot}$}   \\
\multicolumn{1}{c}{}                &
\multicolumn{1}{c}{}                &
\multicolumn{1}{c}{(K)}             &  
\multicolumn{1}{c}{(dex)}           &  
\multicolumn{1}{c}{}                &  
\multicolumn{1}{c}{}                &  
\multicolumn{1}{c}{}                &  
\multicolumn{1}{l}{(d)}            \\

\hline
\\
   3430868 &  8.13 & 4729 & 4.584 & -0.5864 & 0.203 &  6 &  -     \\
   4554830 & 10.33 & 5317 & 4.314 & -0.0051 & 0.039 &  1 & 14.602 \\
   5108214 &  7.83 & 5663 & 3.857 &  0.6404 & 0.028 &  8 &  -     \\
   6442183 &  8.52 & 5760 & 4.006 &  0.4606 & 0.015 & 15 &  -     \\
   7206837 &  9.76 & 6100 & 4.148 &  0.4527 & 0.032 & 35 &  4.050 \\
   7940546 &  7.39 & 5987 & 4.170 &  0.3916 & 0.041 & 37 &  -     \\
   7944142 &  7.81 & 4630 & 2.796 &  1.3908 & 0.121 &  3 &  1.729 \\
  12307366 & 11.50 & 4958 & 3.654 &  0.6328 & 0.006 &  6 &   -    \\
\\
\hline
\end{tabular}
\end{center}
\end{table}

\citet{Karoff2014} calculated the pre- and post-flare acoustic spectra from 
substrings of different lengths before and after the flares. The photometric 
variability associated with the flares was evaluated by measuring the total 
energy in the high-frequency part of both the pre- and post-flare acoustic 
spectra.  Measuring the total power in a given frequency range may not be the most
efficient method of detecting changes in the solar-like oscillations because
the spectra are noise limited.  In other words, a significant part of the
total power comes from the noise and not the actual oscillations.  Instead,
it might be best to compare the amplitudes of individual peaks in the 
periodogram of the global oscillations before and after the flare.  In this 
way one may hope to detect possible systematic increases or decreases of 
amplitude in individual modes as might be expected.

\begin{figure}
\centering
\includegraphics{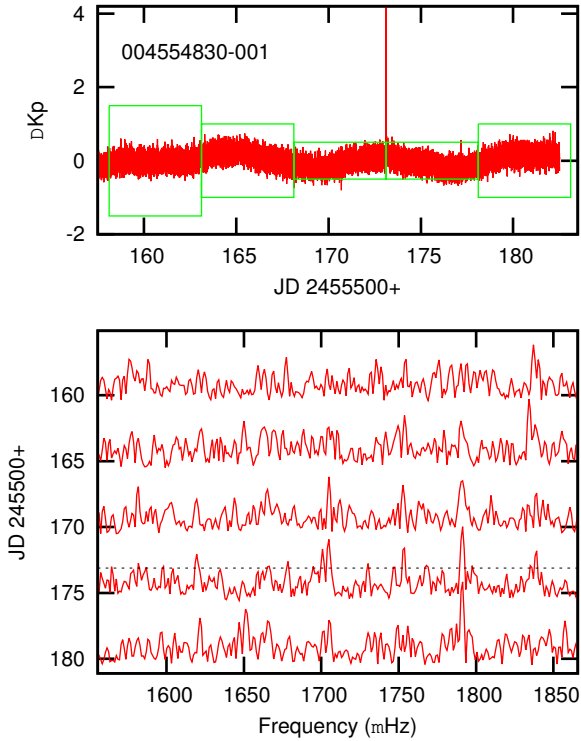}
\caption{Top panel: light curve of KIC\,4554830 showing flare and regions
where periodograms were calculated.  Bottom panel:  periodograms of the
regions shown in the top panel showing the solar-like oscillations.  The
dotted line shows the time of the flare.}
\label{004554830-001}
\end{figure}

In Fig.\,\ref{004554830-001} we show the light curve and periodogram of
KIC\,4554830 around the time of a flare.  The periodogram is calculated using 
two or three 5-d data windows before and after the flare.  Because of the
stochastic nature of solar-like oscillations, changes in amplitude of
individual peaks are to be expected.  One therefore needs to examine
several different frequency peaks and to determine if the changes in
amplitude (increase or decrease) is significantly larger than would 
normally occur.  In KIC\,4554830 the amplitude of the mode at 
1790.99\,$\mu$Hz appears to increase after the flare.  On the other 
hand, the amplitude of the mode at 1839.41\,$\mu$Hz seems to decrease.    
The data windows are independent (no overlap) so one can judge the 
significance of the amplitude changes.  

Similar diagrams may be created for all flare stars with solar-like 
oscillations, but this may not be the best method of detecting starquakes. 
Since we are looking for changes in amplitude of individual modes, we
selected several peaks of large amplitude and fitted a truncated Fourier 
series to the data window using these frequencies.  This gives us the
amplitude and its standard deviation for each frequency peak.  We performed
this calculation in two data windows before and after the flare.

\begin{figure}
\centering
\includegraphics{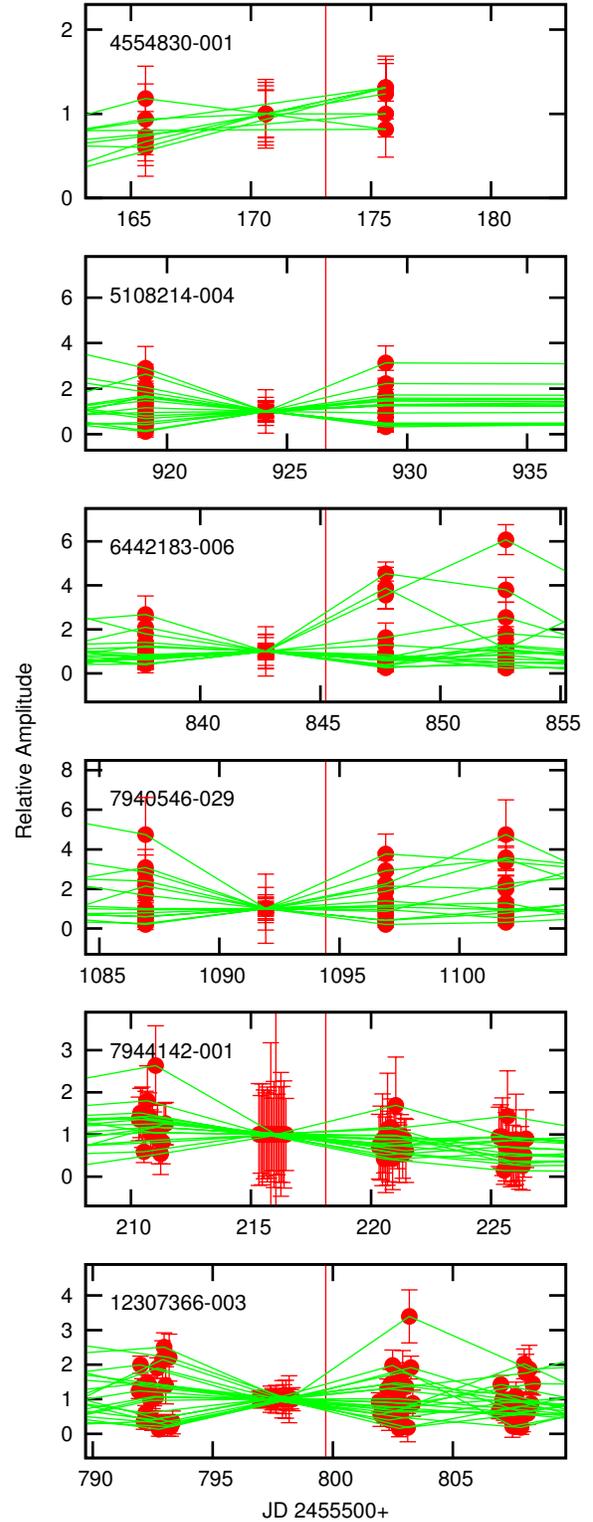}
\caption{Pulsation amplitudes for solar-like oscillations as a function of 
time.  The amplitudes have been normalized to the values just before the 
flare and one-$\sigma$ error bars are shown.  Solid lines join individual
mode frequencies.  The time of the flare is indicated by the vertical line.}
\label{ampsol}
\end{figure}

Fig.\,\ref{ampsol} shows the resulting amplitudes in each 5-d window as a 
function of time for all solar-like modes of sufficient amplitude.  For 
clarity, all amplitudes are normalized to their values in the window just 
before the flare.  The same mode is connected with a solid line.
Judging from the error bars, it is evident that in no case is there a 
clear amplitude change after the flare relative to the pre-flare amplitude.  
We conclude that a superflare has no systematic influence on the amplitudes 
of the solar-like oscillations detectable by this technique. 

This does not mean that starquakes do not occur, of course.  The impulse 
generated by a superflare is bound to cause acoustic disturbances which 
could affect the amplitudes of global modes.  However, the small spatial
scale (high azimuthal degree) of such oscillations lead to cancellation
effects, resulting in whole-disk light variations that are too small to
detect.

\section{Flares in $\delta$~Scuti and $\gamma$~Doradus stars}

\begin{figure}
\centering
\includegraphics[angle=-90]{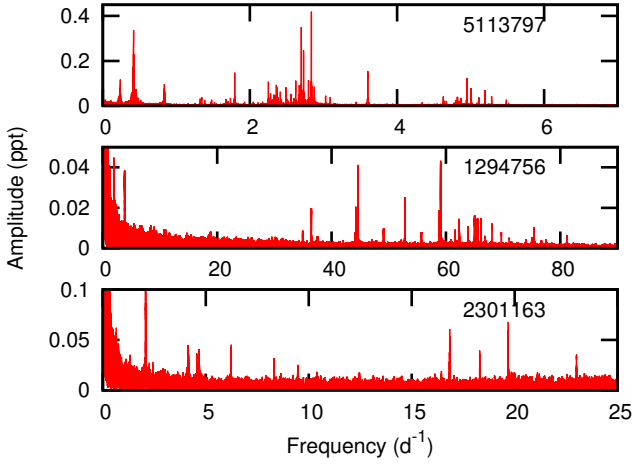}
\caption{Periodograms of a $\gamma$~Dor star (KIC\,5113797, top panel) and
two $\delta$~Sct stars that show flares.}
\label{dsct}
\end{figure}

It is very difficult to detect flares in classical pulsating stars because
the light variations due to the pulsation tend to mask short-lived rapid
excursions such as a flare.  This is especially true for short-period
variables.  Nevertheless, among the SC observations, flares are seen in the 
$\delta$~Sct stars KIC\,1294756 and  KIC\,2301163.  Both these stars have 
extremely low pulsation amplitudes, rendering the flares more easily visible.  
The flare star KIC\,5113797 seems to be a $\gamma$~Dor variable of low 
amplitude.  The periodograms of these stars are  shown in Fig.\,\ref{dsct}.

\begin{figure}
\centering
\includegraphics{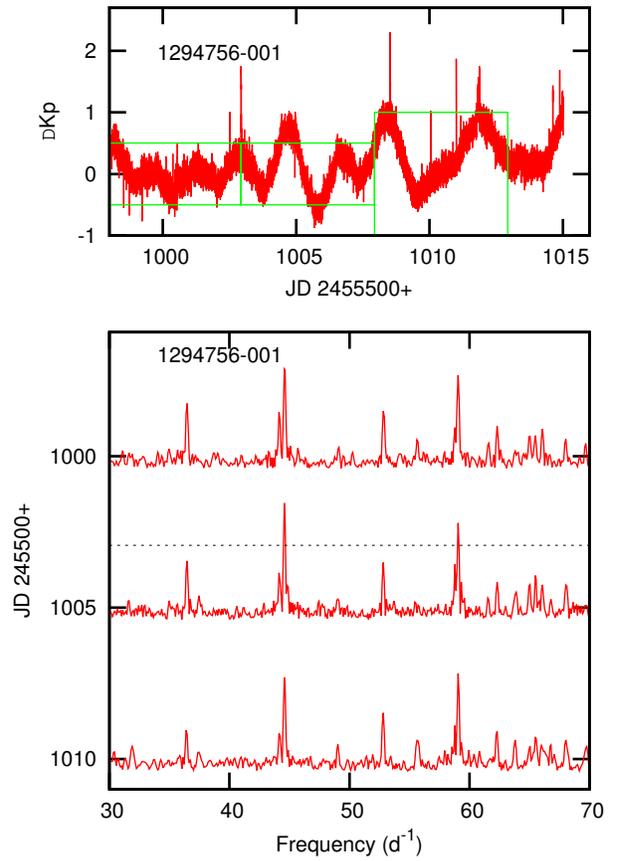}
\caption{Top panel: light curve of the $\delta$~Sct star KIC\,1294756 showing 
flare and regions where periodograms were calculated.  Bottom panel:  
periodograms of the regions shown in the top panel showing the $\delta$~Sct
pulsations.}
\label{001294756-001}
\end{figure}

The oscillations in $\delta$~Sct and $\gamma$~Dor stars are self-excited
with reasonably stable amplitudes and phases, unlike the stochastic
solar-like oscillations.  Hence changes in pulsation amplitude after a
flare might be easier to detect.  Unfortunately, the S/N ratio in the 
$\delta$~Sct pulsations of KIC\,2301163 is too low for the oscillations to be 
detected in a 5-d data window.  Since the effect of a flare impulse on the 
oscillations is expected to dissipate rather quickly, one needs to use a short
data window to optimize its detection.  For this reason KIC\,2301163 was 
excluded from the analysis.  The low oscillation frequencies in the 
$\gamma$~Dor variable  KIC\,5113797 means that very few pulsation cycles can 
be obtained during the 5-d window.  As a result, the amplitudes and phases 
have large errors.  This star, too, was omitted.

In Fig.\,\ref{001294756-001}, part of the light curve of the $\delta$~Sct star
KIC\,1294756 is shown centered on one of the flares.  The boxes show the data 
windows used to construct the periodograms.  Some amplitude changes seem to
occur, particularly for the mode at 34.9317\,d$^{-1}$ which appears to 
decrease after the flare.

\begin{figure}
\centering
\includegraphics{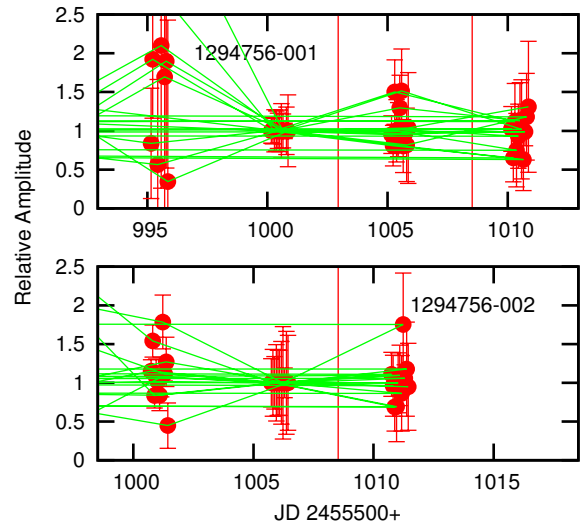}
\caption{Relative pulsation amplitudes for the $\delta$~Sct star KIC\,1294756 
as a function of time.  The amplitudes have been normalized to the values 
just before the flare and one-$\sigma$ error bars are shown.  The time of the 
flare is indicated by the vertical line.}
\label{ampdsct}
\end{figure}

In Fig.\,\ref{ampdsct} we show how the pulsation amplitudes vary with time
before and after the flare. In this figure the data for two flares are
shown.  As before, the amplitudes are relative to the amplitudes just before 
the flare.  Although there are variations in amplitude for many modes, these 
are within the expected errors and are therefore not significant.  We conclude
that even when the pulsation amplitudes are stable, there is no evidence that 
a flare affects the mode amplitudes in a manner that is detectable.

\section{Conclusion}

In this paper we investigate possible periodic light variations arising from 
stellar flares.  By analogy to the Sun, quasi-periodic pulsations (QPP) may
be a result of MHD forces and other processes operating in the flare
loop.  These oscillations, which are often observed in solar flares, have 
also been detected in other stars and have periods in the range of seconds to
tens of minutes.   The period of the oscillation offers the potential to 
probe the magnetic field strength and temperature in the flare.  However, 
nearly all the light in solar QPP is emitted in lines of highly-ionized 
elements, whereas observations of stellar QPP are essentially white light.  
One cannot therefore directly compare solar and stellar QPP, though the
underlying mechanism might still be the same. 

QPP in some stellar flares observed from the ground \citep{Rodono1974, 
Zhilyaev2000, Mathioudakis2006, Contadakis2010,  Qian2012} have been
interpreted as analogous to flare QPP in the Sun, in spite of the very
different light emission properties described above.  These QPP generally
have very short periods and last for several cycles.  We searched for
short-period QPP using wavelet analysis in all flares with high S/N and
found seven flares in five stars in which such an oscillation seems to be 
present (Figs.\,\ref{fast1}--\ref{fast7}).   The periods do not correlate with
$\nu_{\rm max}$ or any stellar parameter.  These are probably examples of
the same phenomenon seen in the ground-based observations discussed above.
They may not be directly comparable with QPP in solar flares, but perhaps a
similar process may be active.

Oscillations in the light curve could also arise as a result of the impulse 
generated by a flare (a starquake).  In the Sun, these are seen as surface 
waves of short spatial wavelength radiating from the location of the flare.  
There is evidence that the impact also affects the amplitudes of the stochastic 
global oscillations.  For some modes a starquake will lead to an increase 
in amplitude, while for other modes a decrease in amplitude may be 
expected.

A significant fraction of stellar flares show clear structures (bumps) on
the decaying branch of the light curve.  We argue on probability grounds that 
these cannot be a result of almost simultaneous multiple flares.  The bump
may be modeled as rapidly decaying QPP.  We tested the possibility that the
bumps may be highly damped forced global oscillations by measuring the
period using wavelet analysis.  If these are forced global oscillations, one 
might expect the period to correlate with some stellar parameter.  We could
find no correlation with the estimated frequency of maximum amplitude, 
$\nu_{\rm max}$, resulting from solar-like oscillations or any other stellar
parameter.  We conclude that the bump cannot be understood as a forced
global oscillation.

Finally, we attempted to detect forced global oscillations resulting from a
starquake.  These are expected to be visible shortly after a flare and are
expected to modify the amplitudes of individual modes in stars where 
solar-like oscillations are detected.  The best way to look for this effect 
is to measure the amplitudes of individual modes before and after a flare.
There are eight stars in the {\it Kepler} short-cadence observations which
show flares and solar-like oscillations.  Examination of the amplitudes of 
modes before and after a flare showed no obvious indication of significant 
amplitude changes.  We conclude that the effect is too small to be detected
in the {\it Kepler} data.

By their nature, random changes in amplitude are a characteristic of
solar-like oscillations and this may mask an amplitude change resulting from
a flare.  The self-excited oscillations in $\delta$~Scuti and
$\gamma$~Doradus stars lead to generally stable amplitudes.  Therefore these
stars may provide better detection of starquakes.  We analyzed the pulsation
amplitudes of two $\delta$~Sct and one $\gamma$~Dor star before and after a
flare.  Again, we were not able to find evidence of amplitude variations due 
to a flare.

A major problem in our understanding of stellar flares is that there are no
corresponding observations in the Sun.  Apart from the huge disparity in
energy, solar flares emit almost entirely in emission lines of highly-ionized 
elements whereas stellar flares are essentially white light flares.
Furthermore, there are no reported observations of QPP in white-light solar 
flares.  Such QPP may be detected in stellar flares observed in X-rays
\citep{MitraKraev2005}.  The standard flare model suggests that the 
white-light emission in solar and stellar flares is triggered by non-thermal 
electrons which originate in the corona. There is a strong correspondence 
between white-light emission and hard X-ray emission \citep{Hudson2006}.  
Moreover, solar hard X-ray bursts often show a high degree of periodicity 
\citep{Aschwanden1994}.  It is possible that the QPP seen in the few {\it
Kepler} flares may be a result of these beams and their effect in the lower 
atmosphere.   However, it may be difficult to understand white light flares 
in A stars where it is generally assumed that a corona is not present.

Our conclusion is that white-light QPP may possibly be seen in some 
{\it Kepler} flare stars, but their nature may differ from QPP in solar flares, although the 
processes involved could be similar.  It would be important to synthesize whole-disk white light 
observations of solar flares.  In this way we may hope to extend what we know 
of solar flares to stellar flares and thereby create a fuller understanding of
the mechanisms involved in solar and stellar flares.

\section*{Acknowledgments} 

This paper includes data collected by the {\it Kepler} mission. Funding for the
{\it Kepler} mission is provided by the NASA Science Mission directorate.
The authors wish to thank the {\it Kepler} team for their generosity in
allowing the data to be released and for their outstanding efforts which have
made these results possible.  

Much of the data presented in this paper were obtained from the
Mikulski Archive for Space Telescopes (MAST). STScI is operated by the
Association of Universities for Research in Astronomy, Inc., under NASA
contract NAS5-26555. Support for MAST for non-HST data is provided by the
NASA Office of Space Science via grant NNX09AF08G and by other grants and
contracts.
 
LAB wishes to thank the South African Astronomical Observatory and the
National Research Foundation for financial support.  TV was sponsored by an 
Odysseus grant of the FWO Vlaanderen and performed in the context of the 
IAP P7/08 CHARM (Belspo) and the GOA-2015-014 (KU~Leuven).  This research was 
also sponsored by the European Research Council research project 
321141 \textit{SeismoSun} (CP, VMN) and STFC consolidated grant ST/L000733/1 
(VMN).  AGK thanks NASA for support (grant NNX14AB68G).  AMB thanks the 
Institute of Advanced Study, University of Warwick for their financial 
support.

\bibliographystyle{mn2e}
\bibliography{qpp}

\label{lastpage}

\end{document}